\documentclass[%
 aip,
 jmp,%
 amsmath,amssymb,
 reprint,%
]{revtex4-1}

\usepackage{graphicx}
\usepackage{dcolumn}
\usepackage{bm}
\usepackage{bm}
\usepackage{graphicx}
\usepackage{color}
\usepackage{amsfonts}
\usepackage{comment}
\newcommand{\be}{\begin{equation}}
\newcommand{\ee}{\end{equation}}
\newcommand{\ba}{\begin{eqnarray}}
\newcommand{\ea}{\end{eqnarray}}
\newcommand{\bd}{\begin{displaymath}}
\newcommand{\ed}{\end{displaymath}}
\newcommand{\nnb}{\nonumber \\}

\begin{document}


\title[Radiation dominated implosion with nano-plasmonics]
{Radiation dominated implosion with nano-plasmonics}

\author{L. P. Csernai}
 \email{Laszlo.Csernai@uib.no}
 \affiliation{Dept. of Physics and Technology, Univ. of Bergen, Norway}
  \affiliation{Sustainability Center, Institute of Advanced Studies, K{\H o}szeg, Hungary}
\author{N. Kroo}%
\affiliation{ 
Hungarian Academy of Sciences, Budapest, Hungary
}%

\author{I. Papp}
\affiliation{Sustainability Center, Institute of Advanced Studies, K{\H o}szeg, Hungary}
\affiliation{%
Dept. of Physics, Babes-Bolyai University, Cluj, Romania
}%

\date{\today}

\begin{abstract}
Inertial Confinement Fusion is a promising option to 
provide massive, clean, and affordable energy for 
mankind in the future.  The present status of research 
and development is hindered by hydrodynamical instabilities 
occurring at the intense compression of the target fuel by
energetic laser beams.

A recent patent combines advances in two fields:
detonations in relativistic fluid dynamics and 
radiative energy deposition by plasmonic nano-shells.

The compression of the target pellet can be moderate
and rapid volume ignition is achieved by a laser pulse, which is 
as short as the penetration time of the light across the pellet. 
The reflectivity of the target can be made negligible, and
the absorptivity can be increased by one or two orders
of magnitude by  plasmonic nano-shells embedded in the target fuel.
Thus, higher ignition temperature can be achieved with 
modest compression.
The short light pulse can heat the target so that most of the 
interior will reach the ignition temperature simultaneously.
This makes the development of any kind of instability impossible, which
would prevent complete ignition of the target.
\end{abstract}

\maketitle

\section{Introduction}

Inertial Confinement Fusion (ICF) is an ongoing activity aiming
for ignition of small pellets of thermonuclear, deuterium-tritium (DT) 
fuel by high-power lasers. The main direction of activity aims for 
strong compression of the fuel, where the resulting adiabatic 
heating would ignite the fuel. The pulse and the compression
should be large and strong enough to keep the compressed fuel 
together for sufficient time for ignition, due to the inertia
of the compressed pellet.
In the present work we present a patented idea how to achieve 
simultaneous volume ignition in the majority of the target
\cite{Patent}.

In the pellet the fusion reaction, 
$D+T \rightarrow n (14.1 {\rm MeV}) +\, ^4\!He (3.5 {\rm MeV})$,
takes place at a temperature of $kT \approx 10$ keV. The produced
$^4\!He$ (or $\alpha$) particles are then deposited in the hot
DT plasma and heat it further. This is the plasma self-heating
(or $\alpha$-heating). The compression wave penetrates into
the plasma with the speed of sound or with the speed of a 
compression shock. There are several facilities with different
configurations attempting to achieve nuclear fusion this way.
A comprehensive summary of these is presented in ref.  \cite{RBOAH16}.

The up to now most successful configuration uses indirect drive.
A spherical pellet of DT fuel of initial outer radius of 1143 $\mu$m
is targeted by laser beams. At the National Ignition 
Facility (NIF) the pellet has a hole in the middle, to reach
better compression, and it has a thin "ablator" layer, which reflects
the incoming light
\cite{Lindl1998,Lindl2004,Haan2011}. 
In this experiment the target
capsule is indirectly ignited by the thermal radiation  coming from the gold
Hohlraum. The Hohlraum is heated by the radiation of 192 laser beams
 \cite{RBOAH16,Nora2015}.
The incoming and reflected light exercised a pressure
and compressed the pellet at NIF, to about $R= 80\ \mu$m 
just before ignition. At this moment the hole in the capsule is already
filled in and the target density was compressed to about 300-700 g/cm$^3$,
\cite{NIF11,RHL16}. Then this target showed the development of Rayleigh-Taylor
instabilities, which reduced the efficiency of ignition.

The initial compression pulse ("low foot")  had lower frequency or
longer wavelength of 100-300 nm, which
therefore had a higher reflectivity on the target, and led to compression.
The reflectivity of light is high ( $> 0.6$ ) for lower frequency light and
decreasing with increasing frequencies. It becomes negligible at 
$\hbar \omega = 1 $ keV. The compression pulse
was followed by a shorter, higher frequency ignition pulse.
The higher frequency pulse has negligible reflectivity and
decreasing absorptivity, having  
$\alpha_K = 10^6$ cm$^{-1}$ at $\hbar \omega = 20 $ eV and
$\alpha_K = 10$ cm$^{-1}$ at $\hbar \omega = 1 $ keV.
If we take a higher frequency, shorter wavelength
of 20 nm in the X-ray range then the absorptivity of the DT fuel is
about $\alpha_K = 10^4$ cm$^{-1}$.\cite{HuCoGo14} 
This means that the full pulse energy 
is absorbed in $10^{-4}$ cm $= 1 \mu$m. That is in a thin surface layer.
The internal domain is heated up due to adiabatic compression, up
to ignition, but the major part of the approximately 10 $\mu$m thin 
compressed surface layer remains cold and only 1 $\mu$m is heated 
up at the outside surface. See Fig. 8 of Ref.~\cite{HuCoGo14}.

\section{Considerations for the target}

In the following consideration we could take (a) a compressed smaller
initial state of radius $R= 80\ \mu$m, which is then not compressed further but 
heated up further with a short penetrating light pulse. Alternatively
we could (b) consider a solid ball of the same amount of DT fuel, which
is then made transparent and ignited by a laser pulse without
significant compression, of radius $R= 640\ \mu$m\,.  
In this second case due to the smaller density we will need a more 
energetic short pulse but also 8 times longer because of the larger size.

Reference \cite{LCJ16} used a similar size target, with an outer ablator 
layer and initial compression. However, they used a special 
cone-in-shell configuration of the target. Through doping the target 
with Cu, they were able to project the K-shell radiation of the target 
when it was radiated by an ultraviolet driver beam. From their images 
in Figure 2 d-f  of \cite{LCJ16}, we see that  ignition is achieved in an 
area of approximately 50 $\mu$m  radius from the center of the target. 
By using a high-contrast laser and a 40 $\mu$m cone tip they 
were able to increase the fast electron coupling to the core from
$<5\%$ to 10-15\% by an increase in the core-density and decrease in 
the source-to-core distance. If this laser-to-electron conversion 
efficiency would be further increased, the total laser energy coupled 
to the core would also increase above 15\%, which may be good if we 
want to achieve fast-ignition inertial-confinement-fusion. 

Here we consider another configuration, without ablator layer
and without too much pre-compression, using the early examples in Refs.
\cite{C87,CS14}.
Then we have a more dilute target
fuel with about 640 $\mu$m radius, similar to case (b). 
With a deuterium-tritium ice as fuel, 
the target density is taken to 1.062 g/cm$^3$. We could increase this
initial density, but only moderately to avoid the occurrence RT
instability. This more dilute target a priory
has smaller absorptivity. If we want to absorb the whole energy
of the incoming laser light on $\sim$ 1.3 mm length, we need an
absorptivity of  $\alpha_K \approx 8 $ cm$^{-1}$. This is about the 
absorptivity of DT fuel for soft X-ray radiation of 1 nm wavelength.
Longer wavelength radiation would have a larger absorptivity, and
would be absorbed in the outside layers of the pellet.

\section{Simplified model and its evaluation}

Consider a spherical piece of matter (E), which is sufficiently
transparent for radiation.  The absorptivity of the  target matter is 
considered to be constant, such that the total energy of
the incoming light is observed fully when the light reaches the
opposite edge of the spherical target. 
 This matter undergoes an exothermic
reaction if its temperature exceeds $T_c$.

\begin{figure}[h]  
\begin{center}
\resizebox{0.8\columnwidth}{!}
{\includegraphics{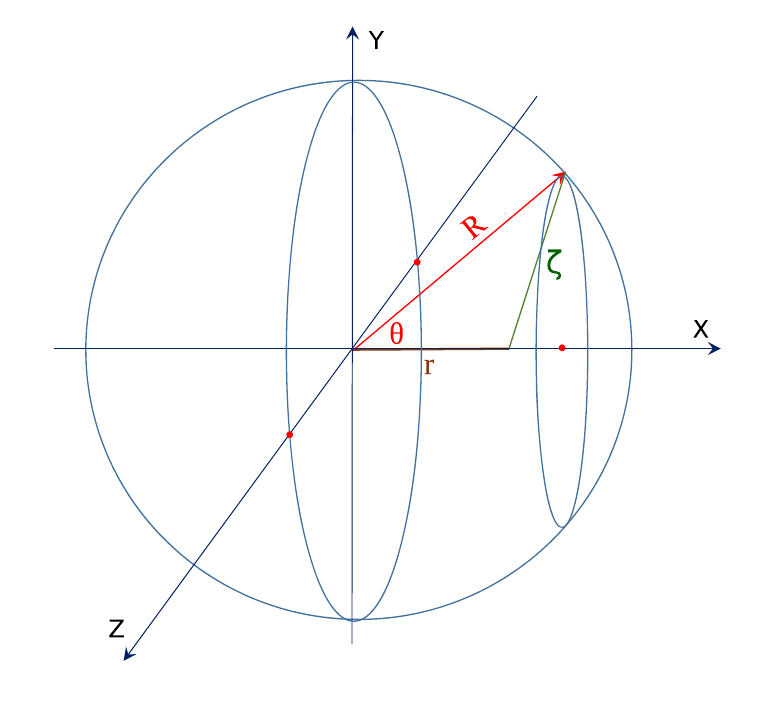}}
\caption{
(color online) 
The sphere of the fuel, with an internal point at radius $r$.
Let us chose the $x$-axis so that it passes through the point at
$r$ and the center of the sphere. Then let us chose a point on the
sphere, and the angle of this point from the $x$-axis is denoted 
by $\Theta$. Then the length between this surface point and the
internal point at $r$ is $\zeta = ( R^2 + r^2  -2 R r \cos \Theta)^{1/2}$. 
The propagation time from the
surface point to the point at $r$ equals $\tau = \zeta/$c.
}
\label{F-1}
\end{center}
\end{figure}

The target matter is surrounded by a set of spherically distributed 
laser beams, 
which emit the radiation necessary to heat up E. Here, for simplicity,
we are neglecting the expansion of the outer shell inwards as well as the 
expansion of the core, so that the core radius $R$ is taken to be constant.
We will measure the length in units of $\mu$m, and the time in units of 
$\mu$m/c.

We intend to calculate the temperature distribution, $T(t,r)$, within the 
sphere, as a function of time, $t$, and the radial distance from
the center of the sphere, i.e. radius $r$.  We have two steps of the
evaluation: \\
(i) In the 1st step we calculate how much energy can reach a given point 
at $r$ from the outside surface of the sphere. Here we have to take into 
account that the outside thermal radiation starts at time $t = 0$, 
so there is no radiation before. The eventual, "Low foot" type 
pre-compression is not included in this dynamical calculation. 
Furthermore, we must consider, which parts of the
outside surface can reach a point inside the
sphere at time $t$, and which are on the backward light-cone of the
point at $r$ and time $t$. 
The integral for the energy density
reaching the point from this part of the two 
dimensional outside surface of the sphere in unit time 
interval, $dt$, is $dU(t,r)/dt$.\\
(ii) Then we have to add up the accumulated radiation at position
$r$, for the previously obtained energy and to obtain the 
time dependence of the temperature distribution, $T(t,r)$, we
have to integrate  $dU(t,r)/dt$ from  $t=0$, for each spatial position.\\
We perform the surface integral of step (i) in terms of integration
for the proper time of the radiation with a delta function, selecting
the surface element, which can reach the given internal point at a time.

Let us study a point within the sphere, at a distance $r$ from
the center. Choose the $x$-axis passing through this point and the
center of the sphere. See Fig. \ref{F-1}.

The surface area of a ring of the sphere at the selected 
polar angle $\Theta$ is $dS = 2 \pi R^2 \sin \Theta \, d\theta$.

Step (i):\\
At a point at $r$ we receive radiation from a layer edge ribbon at 
time $\tau$. The radiation
at distance $\zeta$ is decreasing as $1/\zeta^2$. The total radiation 
reaching point $r$ from the ribbon at $\Theta$ is 
\be
dU(t,r) \propto  \frac{1}{\zeta^2} \delta(\zeta{-}\sqrt{R^2{+}r^2{-}2rR\cos\Theta})\ ,
\ee
where $\tau=\zeta/$c, 
and we should integrate this for the surface of all ribbons.

The average intensity of thermal radiation reaching the surface of 
the pellet amounts to 
$Q$ per unit surface ($\mu$m$^2$) and unit time ($\mu$m/c).
Let us take a typical value for the energy of the total ignition
pulse to be  2\, MJ, in time 10\, ps,  then  \
$Q= 2 {\rm MJ}\ (4 \pi)^{-1} (\cdot 640 \mu {\rm m})^{-2}\ (10 {\rm ps})^{-1}$\  or\ 
$ Q \approx 3.87  \cdot 10^{20}\ {\rm W / cm}^{2} =
1.29 \cdot 10^{10}\    {\rm J\,c / cm}^{3} $.

\begin{figure}[h]  
\begin{center}
\resizebox{0.8\columnwidth}{!}
{\includegraphics{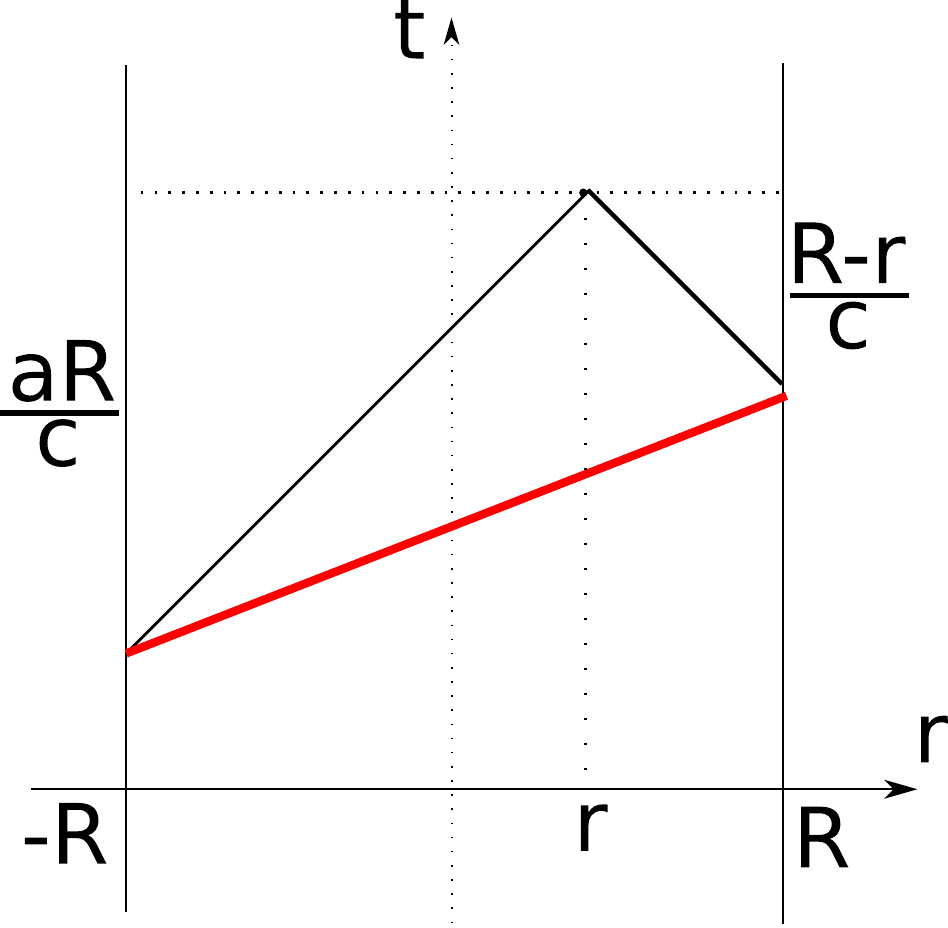}}
\caption{
(color online) 
The boundaries of the integration domains depending on
$r$ and $t$. The domain for the smallest $\tau$ -values
cannot receive radiation, Eq. (2c), because the radiation started at
$(R-r)/$c earlier and it reaches the internal point at $r$ later, 
At the same time the radiation from the opposite 
side reaches the point $r$ also in time $aR/$c. The contour of the
intersection of the backward light-cone with the surface of the 
sphere is indicated with the thick, red line.
When the 
time from the momentum of ignition is longer than
$t=2R$/c the radiation reaches the matter from all
sides at every location $r$. At earlier times the upper boundary
of integration should be evaluated. See Eq. (\ref{E4}).
}
\label{F-2}
\end{center}
\end{figure}

Up to a given time $t$, the light can reach a space-time 
point ($t,r$), inside the 
sphere from different points  of the outside surface, which were emitted in
different times. At early times it may be that none of the surface points are
within the backward light-cone of the point ($t,r$). At later times, 
from part of the
surface points the light can reach ($t,r$),  while at times larger 
than $2R/$c
all internal points can be reached from any surface point of the sphere. Thus,
we calculate first what energy density, $U(t,r)$, we get at a space-time 
point ($t,r$), from earlier
times. At a given point at $r$ measured from the center of the sphere
(assuming that a constant fraction, $\alpha_K$, of the radiation energy is 
absorbed in unit length):
\footnote{
We are using the relation
$\delta[g(x)] = \sum_i [ (1/|g'(a_i)|) \delta(x-a_i)$ where $a_i$-s are the 
roots of $g(x)= \zeta - \sqrt{R^2+r^2-2rRx}$, i.e. $g(a_1)=0$. Now 
$a_1 = (R^2+r^2-\zeta^2)/(2rR)$  and
$g'(x) = rR/\zeta$ so that the integrand is 
$\zeta/(\zeta^2 rR) = 1/ (rR \zeta)$.
The variable $\zeta$ depends on $x$ (or $\Theta$), so we should set the 
integral boundaries in terms of $\zeta$ accordingly.
} 
\ba
&& dU(t,r) = 
\nnb
&&
\alpha_K Q\! \int_0^t \!\!\!\! d\tau\, 
2\pi R^3 \int_0^{\pi} \!\!\!\!d \cos\Theta\,
\frac{\delta(\zeta{-}\sqrt{R^2{+}r^2{-}2rR\cos\Theta})}{R^2+r^2-2rR\cos\Theta}
\nnb
&& = \alpha_K Q\! \int_0^t \!\!\! d\tau\, 
2\pi R^3 \int_{1}^{-1} \!\!\!d x\,
\frac{\delta(\zeta{-}\sqrt{R^2{+}r^2{-}2rRx})}{R^2+r^2-2rRx}
\nnb
&& = 2\pi R^3 \alpha_K Q \cdot (Rr)^{-1} 
\int_{(R{-}r)/c}^{aR/c} \frac{d\tau}{\tau{\rm c}} \ ,
\ea
where the integral over $dx$ gives $1/(Rr\zeta)=(R\,r\,\tau\,c)^{-1}$. 
The time, $d\tau$, integral runs from the nearest point of the backward 
light cone to the surface of the sphere to the furthest point, $aR/c$. 
Here the parameter $a$ will be described later.   See Fig. \ref{F-2}.

Now we introduce a new, dimensionless time variable:
$
q \equiv \tau {\rm c}/R \ .
$
Thus,
\ba
dU(t,r) &=& 
2\pi R^3 \alpha_K Q \cdot (rR{\rm c})^{-1} \int_{1{-}r/R}^{a} \frac{dq}{q}
\nnb
&=&
 2\pi R^2 \alpha_K Q \cdot (r{\rm c})^{-1}\, [ \ln (q) ]_{1{-}r/R}^{a} \ ,
\ea
%
where 
$\zeta{=} \tau$c and
$a$ is the upper boundary of the integral over  the dimensionless time~$dq$.
\be
a = \left\{
\begin{array}{ll}
	1-r/R\ , & q<1{-}r/R \\
	q   \ , & 1{-}r/R<q<1{+}r/R \\
	1+r/R\ , & q>1{+}r/R
\end{array}
\right.
\label{E4}
\ee

Here actually the integral over $dq$ is adding up the contributions
of those surface elements of the sphere, from where radiation reaches the
internal point at $r$ at the same dimensionless time $q$. In the first case the radiation
does not reach the point at $r$ then, in the second part the radiation 
from the closest point of the sphere reaches $r$ but from the opposite point
not yet, in the third case radiation reaches $r$ from all sides.

Thus the energy deposited in unit time at dimensionless time $q$ is
\ba
&&dU(r,q) = \frac{2 \pi R^2 \alpha_K Q}{rc} \times
\nnb
 &&\left\{
\begin{array}{ll}
	\ln[(1+r/R)/(1-r/R)] , & q>1{+}r/R \\
	\ln[q/(1-r/R)] , & 1{-}r/R<q<1{+}r/R \\
	0 , & q<1{-}r/R
\end{array}
\right.
\nnb
\ea
 
Step (ii):\\
Neglecting the compression and assuming constant specific heat $c_v$,
we get that $k_B\,dT = \frac{1}{n\,c_v} dU\cdot dq$, where $k_B$ is the 
Boltzmann constant, and so
\ba
&&k_B\, T(t,r) =  \frac{1}{n\,c_v} \int_0^{t{\rm c}/R} dq \cdot dU(r,q) = 
\frac{2 \pi R^2 \alpha_K Q}{n\,c_v\, r c} \times
\nnb
&&\left\{  
\begin{array}{ll}
\left[q\,\ln\left(1+r/R)/(1-r/R)\right)\right]_{1+r/R}^{tc/R }+ \\
\ \ \ \	(1+r/R)\,\ln[(1+r/R)/(1-r/R)] - 2r/R\, ,\\
\ \ \ \ \ \ \ \ \ \ \ {\rm if:}\ \ \ \ \  tc/R>1{+}r/R\\
\left[q\,\ln\left(q/(1-r/R)\right) -q\right]_{1-r/R}^{tc/R}\, ,  \\
  \ \ \ \ \ \ \ \ \ \ {\rm if:}\ \ \ \ \  1{-}r/R<tc/R<1{+}r/R \\
	0\, , \ \ \ \ \ \ \ {\rm if:}\ \ \ \ \ \ tc/R<1{-}r/R
\end{array}
\right.
\nnb
\ea
 
and so,
\ba
&& k_B\, T(t,r) = H \cdot \frac{R^2}{r} \times
\nnb
&&\left\{  
\begin{array}{ll}
	tc/R\,\ln[(1+r/R)/(1-r/R)] -2r/R\, , \\
\ \ \ \ \ \ \ \ \ \ \ {\rm if:}\ \ \ \ \  \ tc/R>1{+}r/R \\
	tc/R\,\ln[tc/(R/(1-r/R))] -tc/R+1-r/R\, , \\
\ \ \ \ \ \ \ \ \ \ \ {\rm if:}\ \ \ \ \  1{-}r/R<tc/R<1{+}r/R \\
	0\, , \ \ \ \ \ \ \ \ {\rm if:}\ \ \ \ \  \ tc/R<1{-}r/R ,
\end{array}
\right.
\nnb
\ea
where the number density of uncompressed DT ice is \\
$n =  3.045 \cdot 10^{22}$ cm$^{-3}$, and the leading constant, $H$, is
\be
H \equiv \frac{2 \pi  Q}{c \, c_V}\, \frac{\alpha_K(r)}{n} 
= 8.54 \cdot 10^{-12} {\rm J / cm}.
\ee

If the absorptivity is varying, $\alpha_K = \alpha_K(r)$, then it follows:
\be
k_B\,T(t,r)  =
\frac{2 \pi  Q R}{c \, c_V\, n}\, 
 \left\{
\begin{array}{ll}
0 \ ,\ \ \ \ \ {\rm if:}\ \ \ tc < R{-}r \\
\frac{\alpha_K(r)tc}{r} \left( \ln \frac{tc}{R-r} - 1 \right) + \frac{R-r}{r}\ , \\
 \ \ \ \ \ \ \ \ \ {\rm if:}\ \ \  R{-}r < tc < R{+}r \\
\frac{\alpha_K(r)tc}{r} \ln \frac{R+r}{R-r} - 2 \ ,\\
 \ \ \ \ \ \ \ \ \ {\rm if:}\ \ \ \ tc > R{+}r
\end{array}
\right.
\ee

The surface of the discontinuity is characterized by 
the $T(t,r) = T_c$ contour line. If $T_c$ is the ignition temperature, then
here the DT ignition takes place on this contour line in the space-time.
The tangent of this line is if $tc>R+r$ :
\ba
&& {\left( \frac{\partial r}{c\partial t} \right)}_{ T_c} = 
 {\left( \frac{\partial T}{c\partial t} \right)}_{ T_c} \Bigg/ 
 {\left( \frac{\partial T}{\partial r} \right)}_{ T_c} =\\ 
 &&{\ln \frac{R+r}{R-r} }\!\! \Bigg/\!\!\!
 \left\{ tc { { \left\lbrack \frac{2R}{R^2{-}r^2}{+}
 \left(\!\frac{\alpha'_K(r)}{\alpha_K(r)}{-}\frac{1}{r}\!\right) \ln  
 \frac{R+r}{R-r} \right\rbrack}\! } \right\}
\ea
So the point $(t_c,r_c)$ where the space-like and time-like 
parts of the surface meet, from 
${\left( {\partial r}/{(c\partial t)} \right)}_{T_c} = 1$
is :
\begin{equation}
t_c = 
{\left\{\!\frac{2cR}{R^2-r_c^2}\Bigg[ \!\ln 
\frac{R+r_c}{R-r_c}\phantom{^I}\Bigg]^{-1} \!\!\!\!\!\! +
\left(  \frac{\alpha'_K(r)}{\alpha_K(r)}   {-} \frac{c}{r_c} 
\right)\right\} }^{-1}
\end{equation}

This line $t = t_c(r_c)$ separates the Space-like and Time-like branch 
of the discontinuity of $T(t,r) = T_c$.

The discontinuity initiates at $r=R$ and $t=0$ and it propagates 
first slowly inwards. Due to the radiative heat transfer 
the contour line of ignition, $T(t,r) = T_c$,
accelerates inwards, and at $r_c = r_c (T_c)$ it develops smoothly 
from space-like into a time-like discontinuity.

The same type of gradual development from space-like into time-like 
detonation occurs in the last, {\em hadronization}, 
phase of ultra-relativistic heavy ion collisions
\cite{FO-HI}.
If we include radiative heat transfer in these scenarios, 
the transition from space-like 
to time-like deflagration will be gradual. This, however, 
requires more involved numerical calculations.

\begin{figure}[htb]
\begin{center}
\resizebox{0.8\columnwidth}{!}
{\includegraphics{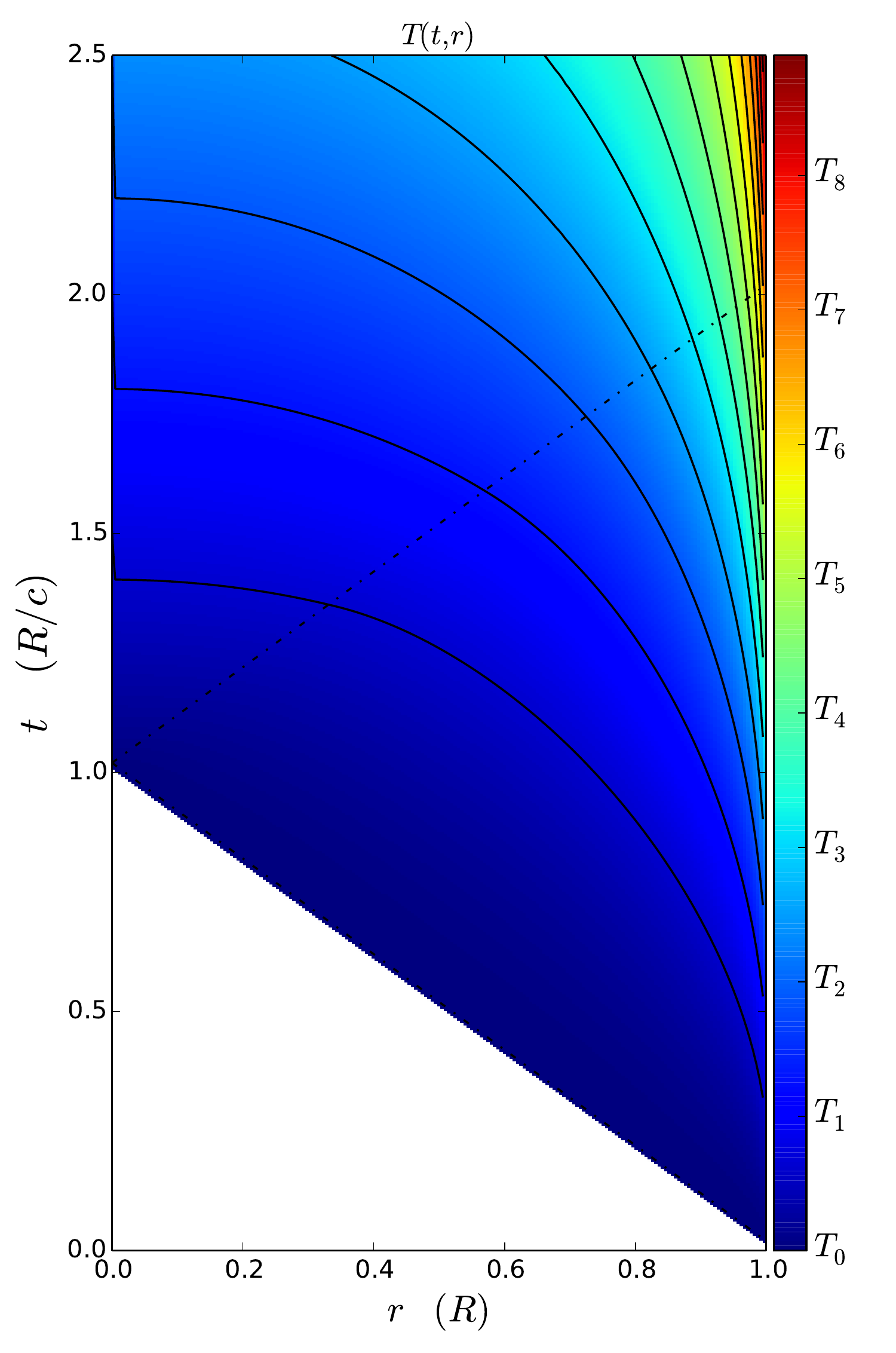}}
\caption{
(color online) 
Analytic solution of the radiation dominated implosion model
of rapid ignition.
The temperature distribution in function of distance and time. 
The lower boundary and the dotted line represent the light cones
from the left and right edge of the pellet, initiated at $t=0$. 
The absorptivity of the pellet is  $\alpha_K = 8 $ cm$^{-1}$.
The temperature is 
measured in units of  
$T_1 = H\cdot R= 21.3\,{\rm keV} $,
and $T_n = n \cdot T_1$.}
\label{F-3}
\end{center}
\end{figure}

In Fig. \ref{F-3} we see the constant temperature contour lines,
$T(t,r)=$const., in the space-time, from the analytical solution.
For the first or second $T=$const. contour line the time-like detonation
region extends from the center of the pellet to the half of radius
R. This is only about 12\% of the volume, so the time-like detonation,
in itself, cannot achieve total simultaneous volume ignition, and 
in the outside region instabilities might develop!

\section{Variable absorptivity}

In order to study the effect of variable absorptivity we reformulated 
the numerical model to perform all integrals of the model numerically.
This will enable us to study the configuration where the 
pellet is manufactured with nano-shells inside, which regulate the 
absorptivity of the DT ice pellet.

\begin{figure}[htb]
\begin{center}
\resizebox{0.8\columnwidth}{!}
{\includegraphics{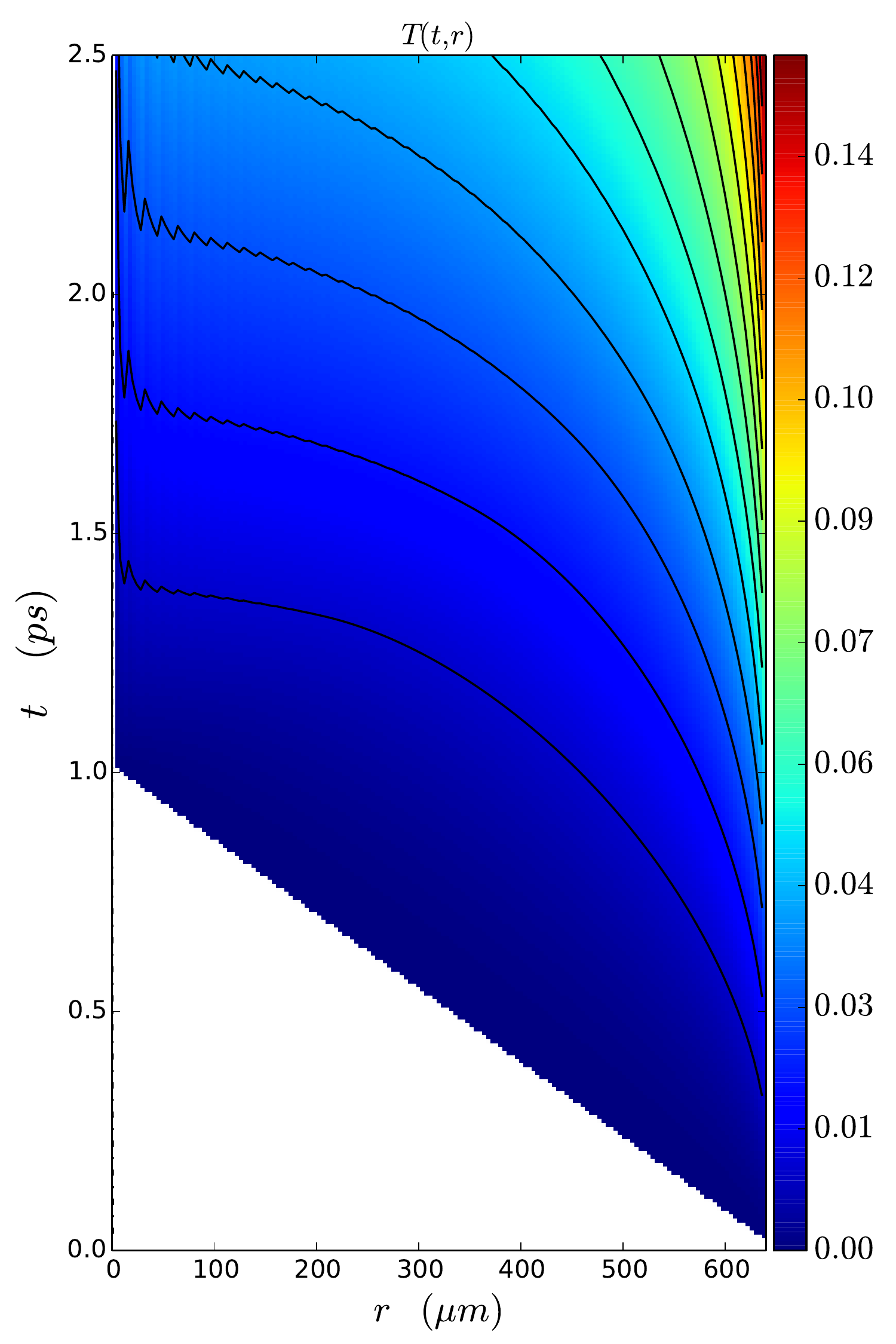}}
\caption{
(color online) 
Numerical solution of the radiation dominated implosion model
of rapid ignition.
The temperature distribution in function of distance and time. 
The dotted line represents the light cone. 
The absorption coefficient is constant 
$\alpha_K = 8 $ cm$^{-1}$.
The decrease of radiation flux due to the absorption is neglected.
The temperature is 
measured in units of  
$T_1 = H\cdot R= 21.3\,{\rm keV} $,
and $T_n = n \cdot T_1$. The finite numerical resolution leads to the
fluctuations near $r = 0$.
The color code for the temperature, $T(t,r)$, is given in units of (MeV).
}
\label{F-4}
\end{center}
\end{figure}

At first we still assumed constant absorptivity with the same value
$\alpha_K = 8 $ cm$^{-1}$.
In Fig. \ref{F-4} 
the numerical solution gives qualitatively identical
result, but at small radii the numerical uncertainty leads to visible
fluctuations and somewhat smaller central temperature increase.
The domain of time-like , i.e. simultaneous ignition takes place 
in the central part of the pellet up to a radius of $r = 370 $ $\mu$m.

In \cite{Hu2014}, Rochester and NIF experimental data were studied
and analysed with opacity data, extracted both from basic principles
and from comparison with ICF experiments. The absorption coefficient
$\alpha_k$ (cm$^{-1}$), defined by $I(x) = \exp(-\alpha_k\,x) I_0$, 
and the Rosseland and Planck opacities, $K_R$ defined
by  $I(x) = \exp(-\alpha_{K_R}\,\rho\,x) I_0$, were estimated and used
to simulate the space-time development of ICF direct ignition 
experiments.

In our previous calculations we used the absorption coefficient
$\alpha_k$ (cm$^{-1}$), and the approximation that the intensity of the
incoming laser light flux is sufficiently large, so that its decrease by the
absorption is negligible. 

With increased absorptivity one could reach more rapid heating. The fusion 
reaction rate per unit volume and unit time is $f = n_D n_T <\sigma v>$,
depends on the D and T densities, $ n_D n_T $, and the reactivity, 
$ <\sigma v> $. Due to the fusion cross section, the reactivity is increasing
up to about $T = 70-100$ keV, and then decreases again. Thus, we could
aim for a heating up to this temperature, with smaller pre-compression.

\section{Variation of absorptivity by Nanotechnology}

Doping inertial confinement fusion pellets with golden nano-sphere
shells enables us to achieve the desired variable absorptivity
\cite{katsuaki2016}.
Nano-shells irradiated by laser light exhibit a resonant
light absorption, which can increase the plasmon field-
strength by up to a factor of 40-100 or more
\cite{N_Halas}.
At present
experimentally realizable nano-shell sizes ranges for core
sizes: 5-500 nm, and for shell thickness: 1-100 nm.
In a fuel target prepared initially with implanted Au
nano-shells, after pre-compression we can have nano-shells
of a radius $\approx 10$ nm.

The resonant light frequency of the nano-shell can be
tuned in a very wide range by changing the 
size and thickness of the nano-shell. 
If the Core $(r_1)$ versus the Shell thickness, 
$r_1/(r_2 - r_1)$ is changed from 2 to 800 the resonant
wavelength changed from $0.5$ to $10 \cdot 10^3$ nm \cite{Loo04}. 
For our
purposes of short wavelength, X-ray photons the smaller
and relatively more thick nano-shells are relevant. An
eventual pre-compression modifies the nano-shells in this
direction.

At the resonant frequencies the
nano-shells are able to absorb resonantly a rather high
portion of incoming light. We can define the $absorption$,
$scattering$ and $extinction$ efficiencies, 
$Q_{abs}$, $Q_{sca}$,  and $Q_{ext}$,
respectively \cite{Lee05}, where these coefficients, $Q_i$, describe how
much part of the energy of the incoming light is 
absorbed or scattered by the nano-shell, compared to its
geometrical cross-section, G, i.e. for a sphere of radius
$R$, $G = R^2\, \pi$.

The nano-shells can be tuned to either larger absorption
efficiency or larger scattering efficiency. For our purposes
the larger absorption efficiency is optimal. The resonance
extinction or absorption efficiency can reach a factor
10 or even more \cite{Alam06}.

Thus, the absorptivity of the target material of the pellet can be
regulated by the density of implanted nano-shells. The target 
DT fuel has a bulk absorption coefficient,
$\alpha_{k0}$ (cm$^{-1}$). 
If we implant nano-shells with cross 
section $G=R^2 \pi$, with a density $\rho$ (cm$^{-3}$) 
then the absorptivity will increase to
\be
\alpha_{k} = \alpha_{k0} + \alpha_{ns}\ , 
\ee
where the absorptivity of nano-shells, $\alpha_{ns}$, is 
\be 
\alpha_{ns} = \rho\ G\ Q_{abs}.
\ee
For a nano-shell of $R=30$nm  the additional contribution would be
$ \rho\ G\ Q_{abs} = \rho\ Q_{abs}\ 0.283$cm$^2$. Consequently
for a typical nano-shell density 
\cite{James07}
of $\rho = 10^{11}/$cm$^3$ and a $Q_{abs} \approx 10$,
we can reach an additional absorptivity of 
\be 
\alpha_{ns} = 28.3\ {\rm cm}^{-1}\ .
\ee
Higher nano-shell density and higher absorption efficiency can also
be achieved. A pre-compression of the target fuel would further
increase the absorptivity.

Absorption for DT densities in the range of 
$\rho = 5 - 200$g/cm$^3$, $T \approx 10^5$ K were obtained 
in the range of 
$$
\alpha_{k0} = 10^{-1} - 10 \ {\rm cm} ^{-1}
$$ 
as the light
frequency increased to  $\hbar\omega = 1-10$keV.  I.e. the typical
light mean free path was about
1-10$^4\mu$m. Thus, while for low frequency radiation, 
$\hbar\omega = 1 - 100$eV 
the DT target is quite opaque, at higher frequencies or energy
it is much more transparent. This leads to the result that the
initial lower energy ("Low foot", "High foot") pulse leads primarily to
compression. This effect is enhanced further with the application of
the thin ablator sheet on the surface of the pellet
\cite{Benredjem2014}.

The additional opacity of nano-shells with typical nano-shell densities
can increase the absorptivity by up to 
$$
\alpha_{ns} = 20-30\ {\rm cm}^{-1}
$$
which makes the fast ignition possibilities very versatile in
this light frequency range. 
We can experiment with  variable absorptivity, which is the 
normal high temperature, high frequency absorptivity of the 
DT fuel, $\alpha_{k0} \approx  1$ cm$^{-1}$ at the outer 
edge of the pellet (i.e. at $R=640\,\mu$m) while in the center,
it is $\alpha_{ns} = 20-30\ {\rm cm}^{-1}$ , (i.e. about up to 30 
times more).  The space time profile of the ignition, will then depend
on the radial profile of the increasing nano-shell doping towards the
center of the pellet. We could optimize this by achieving the largest
simultaneous volume ignition domain, which eliminates the 
possibility of developing instabilities.  

\begin{figure}[ht]
\begin{center}
\resizebox{0.75\columnwidth}{!}
{\includegraphics{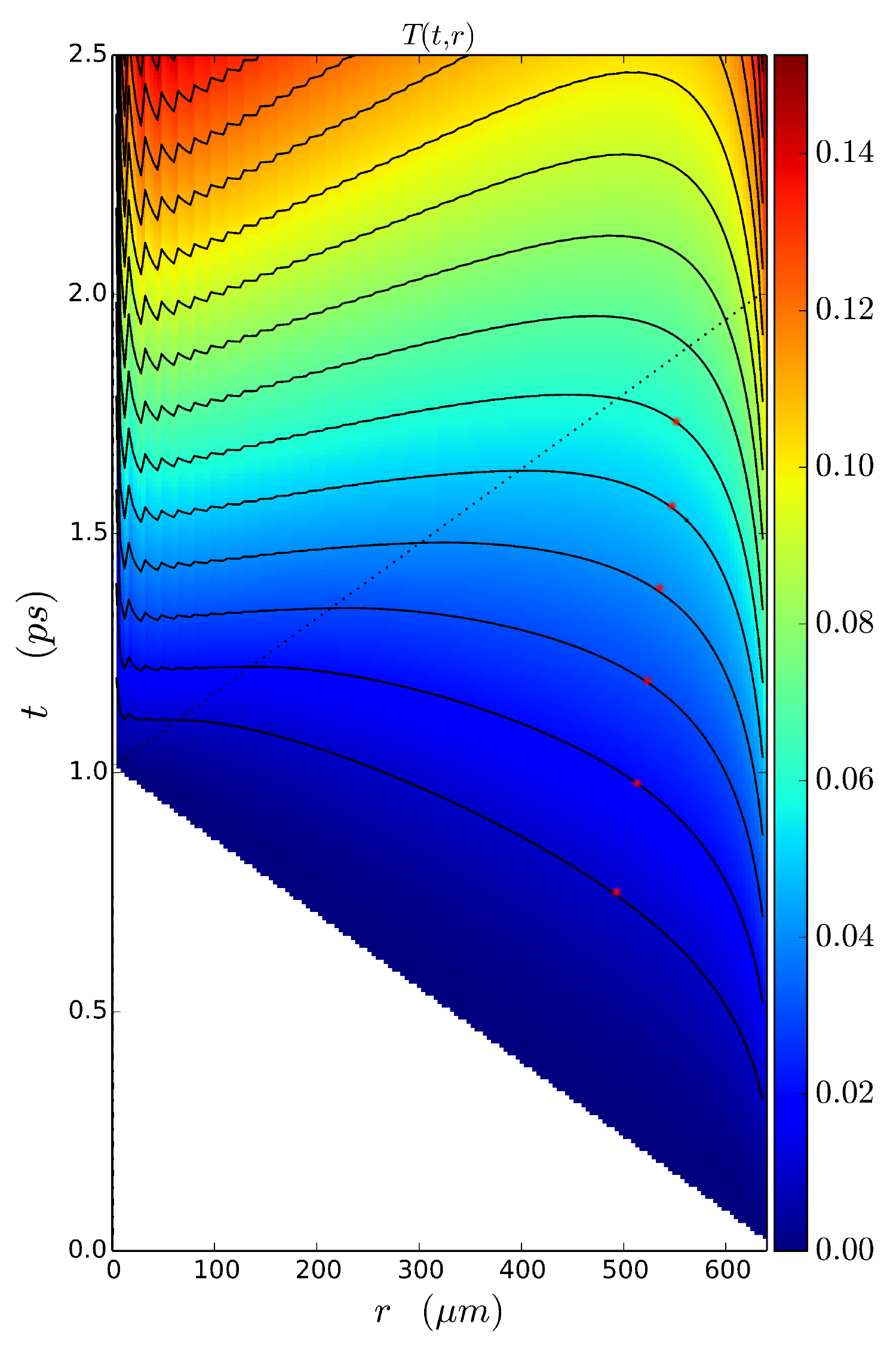}}
\caption{ (color online) 
Numerical solution of the radiation dominated implosion model
of rapid ignition.
The temperature distribution in function of distance and time. 
The dotted lines represent the light cone. 
The absorption coefficient is linearly changing with the radius.
In the center, $r=0$, $\ \alpha_K = 30 $ cm$^{-1}$ while at the 
outside edge $\alpha_K = 8 $ cm$^{-1}$.
The decrease of radiation flux due to the absorption is neglected.
The temperature is measured in units of  
$T_1 = H\cdot R= 21.3\,{\rm keV} $,
and the $T_n(r) = n \cdot T_1=$const. contour lines are shown.
The color code for the temperature, $T(t,r)$, is given in units of (MeV).
The finite numerical resolution leads to the fluctuations near $r = 0$,
this is a numerical artifact.
The stars on the temperature contour lines indicate the transition 
from space-like front at the outside edge to time-like front in the middle.
The points of the middle part are not causally connected, so instabilities 
cannot develop.
}
\label{F-5}
\end{center}
\end{figure}

\section{Conclusions and discussions}

Using nano-technology for ICF is mentioned recently
\cite{Nano-Rods}.  By placing aligned nano-rods or nano-wires
on the surface of the pellet and irradiating irradiating it with 
femtosecond laser pulses of relativistic intensity, leads to a 
plasma with peak electron intensity and pressure. However, this
pressure would lead to a pressure driven adiabatic compression and
heating, which can lead to Rayleigh-Taylor Instabilities, preventing
simultaneous volume ignition.

In this model estimate, we have neglected the compression of the 
target solid fuel ball, as well as the reflectivity of the target matter.   
The relatively small absorptivity made it possible that the radiation
could penetrate the whole target.  With the model parameters we used
the characteristic temperature was $T_1 = 21.3\,{\rm keV} $, which 
is larger than the usually assumed ignition temperature, while 
our target is not compressed so the higher temperatures may be
necessary to reach ignition according to the Lawson criterion.
If we can achieve ignition at somewhat lower temperature than $T_1$,
the ignition surface in the space time includes a substantial
time-like hyper-surface, where instabilities cannot develop, because
neighboring points are not causally connected. 

From looking at the constant temperature contour lines in function of distance 
and time (Fig. \ref{F-5}), we see that the detonation at a 
higher critical temperatures,
$T_c \approx T_4 - T_6$  occurs
when the radiation reaches the matter from the other side also.  
At these contour lines of $T \approx 70-90$ keV, 
about 90\% of the interior ignites at the same 
time, so the ignition is simultaneous for about 73\% of the total volume. 
In this domain no instabilities may occur and the thinner external crust
is not the origin of instabilities according to the experience from 
LLNL.

We can also apply this model is to a
pre-compressed, more dense target, which is transparent and has
larger absorptivity. In this situation the ignition temperature can be
somewhat smaller, but we still can optimize the pulse strength
and pulse length to achieve the fastest complete ignition of the 
target.

We can see if we neglect the importance of the speed of light, 
the theory would be far-fetched from reality. 
It is important to use the proper relativistic treatment
to optimize the fastest, more complete ignition, 
with the least possibility of 
instabilities, which reduce the efficiency of ignition.

\section*{Acknowledgements}

Enlightening discussions with Igor Mishustin, and Horst St\"ocker 
are gratefully acknowledged.
This work is supported in part by the Institute of Advance Studies, 
K{\H o}szeg, Hungary.

\section*{References}


\end{document}